\documentclass[10pt]{article}

\usepackage[final]{epsfig}
\usepackage{amsmath}
\usepackage{parskip}

\def\lsim{\mathrel{\rlap{\lower4pt\hbox{\hskip1pt$\sim$}}
    \raise1pt\hbox{$<$}}}         
\def\gsim{\mathrel{\rlap{\lower4pt\hbox{\hskip1pt$\sim$}}
    \raise1pt\hbox{$>$}}}         

\def\overleftrightarrow#1{\vbox{\ialign{##\crcr
    $\leftrightarrow$\crcr
    \noalign{\kern 1pt\nointerlineskip}
    $\hfil\displaystyle{#1}\hfil$\crcr}}}

\begin{document}

\hspace{12cm}{\bf TUM/T39-03-30}\\

\begin{center}
{\bf Why a long-lived fireball can be compatible with HBT measurements}
\footnote{work supported in part by BMBF and GSI}
\end{center}
\begin{center}
{Thorsten Renk$^{a}$}

{\small \em $^{a}$ Physik Department, Technische Universit\"{a}t M\"{u}nchen,
D-85747 Garching, GERMANY}

\end{center}
\vspace{0.25 in}

\begin{abstract}
The common interpretation of HBT data measured at top SPS energies leads to apparent source lifetimes
of 6--8 fm/c and emission duration of approximately 2--3 fm/c.
We investigate a scenario with continuous pion emission from a
long-lived ($\sim$ 16 fm/c)  thermalized source in order to show
that it is not excluded by the data.
Starting from a description of the source's spacetime expansion based 
on gross thermodynamical properties of hot matter
(which is able to describe a number of experimental observables), 
we introduce the pion emission function with a contribution
from continuous emission during the source's lifetime 
and another contribution from the final breakup and proceed by calculating the HBT
parameters $R_{out}$ and $R_{side}$. The results are compared
with experimental data measured at SPS for 158 AGeV central Pb-Pb collisions.
We achieve agreement with data, provided that some minor modifications
of the fireball evolution scenario are made and find that the parameter
$R_{out}$ is not sensitive to the fireball lifetime, but only
to the duration of the final breakup, in spite of the fact that emission takes place
throughout the whole lifetime. We explicitly demonstrate
that those findings do not alter previous results obtained within this framework.
\end{abstract}

\vspace {0.25 in}

\section{Introduction}
\label{sec_introduction}

One of the main goals in the study of ultrarelativistic heavy-ion
collisions (URHIC) is to understand the conditions realized
in the center of such a collision. If the created system is sufficiently
thermalized, one can in principle link measurements with information
coming from lattice simulations about the equation of state (EoS) of hot QCD matter
(see e.g. \cite{Karsch}).
However, in order to make use of the EoS, some information on
the spatial expansion pattern of the matter is extremely helpful.

HBT correlation measurements can provide at
least part of this information (see \cite{HBTReport, HBTBoris} for a review).
However, such measurements can only reveal the correlation lengths
of the source, which result from a complex interplay of temperature,
flow pattern and true geometrical shape.
Commonly, simple (often Gaussian) parameterizations of the emission source
are used to interpret the experimental information and extract the
geometrical (Gaussian) radius of the source $R_G$, its transverse expansion velocity
$v_\perp$, from the longitudinal correlation radius $R_{long}$ the source lifetime
$\tau_f$ and from $R_{out}/R_{side}$ its emission duration $\Delta\tau$.
 In  \cite{CERES-HBT} (discussing HBT measurements of Pb-Pb collisions at SPS),
these parameters are  found to be
$R_G\approx 6$ fm,  $\tau_f \approx 6-8$ fm/$c$, $v_\perp \approx 0.5c$ and
$\Delta\tau \approx 2-3$ fm/$c$ for an assumed
temperature at breakup of $T_f = 120$ MeV. These numbers point to a rather rapidly
expanding and decaying system.

However, there is a subtle problem with this interpretation:
If the system expands from a root mean sqare radius $R_{rms}$ of about $4.5$
fm (as found by a calculation of the overlap of the colliding nuclei)
to the measured $R_{rms}$ of about $8.5$ fm (assuming $R_G = 6 $ fm), while
at the same time the fireball front (for a Gaussian distribution, take e.g. $R_{rms}$
for the front) expands to a velocity of $\sim 0.5c$, we may expect the
following equations to approximate the problem \cite{FREEZE-OUT}:
\begin{equation}
\label{E-tside}
R_{rms}(\tau_f) - R_{rms}(0) = \frac{a}{2} \tau_f^2 \qquad \text{and} \qquad
v_\perp(\tau_f) = a \cdot \tau_f
\end{equation}
Solving for $a$ and $\tau_f$ we find $\tau_f \approx 15$ fm/c, clearly incompatible
with the lifetime extracted from $R_{long}$. What is wrong here?

In \cite{FREEZE-OUT}, it has been pointed out that the relation \cite{Sinyukov}
\begin{equation}
\label{E-Sinyukov}
R_{long} = \tau_f(T_f/m_t)^{1/2}
\end{equation}
(with transverse mass $m_t = \sqrt{p_t^2 + m^2}$ and $p_t$ the transverse
momentum of a particle with mass $m$)
linking the measured $R_{long}$ and $\tau_f$ is based on a backwards
extrapolation of the measured freeze-out state, assuming that the
longitudinal expansion takes place with the same velocity at all times before $\tau_f$.
However, this relation ceases to hold if the system undergoes
accelerated longitudinal expansion also (in this case, the
lifetime can be significantly longer than extracted from
(\ref{E-Sinyukov})), so the estimate done from the transverse dynamics
is probably more reliable.

This, however, immediately gives rise to the following question: If a 
thermal source expands for $O(15)$ fm/c, how can it be that its apparent emission
duration is only $O(2)$ fm/c? In order to answer this question, 
one needs to specify the space-time evolution of the source beyond
the schematic parameterization used in \cite{CERES-HBT}. We will employ
a description of this evolution based on bulk thermodynamic properties
of the fireball to investigate this question.

The paper is organized as follows: After an introduction to
the fireball expansion model, we discuss in greater detail
why longitudinal acceleration helps to match the timescales
extracted from $R_{long}$ and $R_{side}$. Then we specify how 
this model is used to calculate the relevant HBT parameters. We subsequently compare
the results to the measured data and explicitly demonstrate the
differences to the standard parameterizations. Then we
discuss the modifications to our scenario which are required to
obtain good agreement with the data. Finally we investigate the
physics behind the observed long fireball lifetime and the small
difference between $R_{out}$ and $R_{side}$ and discuss the
impact of the modifications introduced before to previous results
obtained within the same framework.

\section{Fireball evolution and hadron emission}

In this section, we briefly outline the general framework of the
fireball evolution model which we use in the following as a starting point 
for the calculation of hadron emission and ultimately the
calculation of HBT radius parameters. 

\subsection{Expansion and flow}

Our fundamental assumption is to treat the fireball matter as thermalized
from an initial proper time scale $\tau_0$ until breakup time $\tau_f$. For simplicity,
we assume a spatially homogeneous distribution of matter. Since some volume elements 
move with relativistic velocities, it is sensible
to choose volumes corresponding to a given proper time $\tau$ for the calculation of
thermodynamics, hence the thermodynamic parameters temperature $T$, entropy density
$s$, pressure $p$, chemical potentials $\mu_i$ and energy density $\epsilon$ become
functions of $\tau$ only for such a system. In the following, we refer to $\tau$ as the
time measured in a frame co-moving with a given volume element.

In order to make use of the information coming from lattice QCD
calculations, we proceed by calculating the thermodynamical response to
a volume expansion that is parametrized in such a way as to reproduce the experimental
information about the flow pattern and HBT correlations as closely as possible.

As a further simplification, we assume the volume to be cylindrically symmetric around
the beam (z)-axis. Thus, the volume is characterized by the longitudinal extension
$L(\tau)$ and the transverse radius $R(\tau)$ and we find
\begin{equation}
\label{E-Volume1}
V(\tau) = \pi L(\tau) R^2(\tau).
\end{equation}

In order to account for collective flow effects, we boost individual volume
elements according to a position-dependent velocity field. For the transverse
flow, we make the ansatz
\begin{equation}
\label{E-FlowProfile}
\eta_T(r, \tau) = r/R_{rms}(\tau)\eta_T^{rms}(\tau)
\end{equation}
where $R_{rms}(\tau)$ denotes the
root mean square radius of the fireball at $\tau$ and $\eta_T^{rms}(\tau)$ the
transverse rapidity at $R_{rms}$.

For the longitudinal dynamics, we start with the experimentally measured width of the rapidity 
interval of observed hadrons $2\eta_f^{front}$ at breakup. From this, we compute the longitudinal velocity of the
fireball front at kinetic freeze-out
$v_f^{front}$. We do not require the initial expansion velocity $v_0^{front}$ to coincide
with $v_f^{front}$ but instead allow for a longitudinally accelerated expansion. 
This implies that during the evolution $\eta = \eta_s$ is not valid (with $\eta = \text{atanh }v$ and
$\eta_s$ the spacetime 
rapidity $\eta_s = 1/2 \ln (t+z)/(t-z)$) in contrast to the non-accelerated case.

The requirement that the acceleration should be a function of $\tau$ and
that the system stays spatially homogeneous for all $\tau$ determines
the velocity field uniquely once the motion of the front is specified.
We solve the resulting equations numerically \cite{Synopsis}
and find that for not too large rapidities $\eta < 4$ and accelerations
volume elements approximately fall on curves $const. = \sqrt{t^2 - z^2}$ 
and that the flow pattern can be approximated
by a linear relationship between rapidity $\eta$ and  spacetime rapidity $\eta_s$ as
 $\eta(\eta_s) = \zeta\eta_s$
where $\zeta = \eta^{front}/\eta_s^{front}$
and $\eta^{front}$ is the rapidity of the cylinder front. 
In this case, the longitudinal
extension can be found calculating the invariant volume $V = \int d\sigma_\mu u^\mu$ as
\begin{equation}
L(\tau) \approx 2 \tau \frac{\text{sinh }((\zeta -1) \eta_s^{front}(\tau))}{(\zeta -1)} 
\end{equation}
with $\eta_s^{front}(\tau)$
the spacetime rapidity
of the cylinder front. This is an approximate
generalization of the boost-invariant relation $L(\tau) = 2 \eta^{front} \tau$ which can be derived
for non-accelerated motion. A more detailed description of the expansion can be
found in \cite{Synopsis}.

\subsection{Parameters of the expansion}

In order to proceed, we have to specify the longitudinal acceleration $a_z(\tau)$ at the fireball front 
(which in turn is used to calculate $\eta_s^{front}(\tau)$ numerically), the
initial front velocity $v_0^{front}$ and the expansion of the radius $R(\tau)$
in proper time.

We make the ansatz
\begin{equation}
a_z = c_z \cdot \frac{p(\tau)}{\epsilon(\tau)}
\end{equation}
which allows a soft point in the EoS where the ratio $p/\epsilon$ gets small
to influence the acceleration pattern. $c_z$ and $v_0^{front}$ are 
model parameters governing the longitudinal expansion and fit to data.

Since typically longitudinal expansion is characterized by larger
velocities than transverse expansion, i.e. $v_z^{front} \gg v_T^{front}$, we
treat the radial expansion non-relativistically. We assume that the radius of the
cylinder can be written as 
\begin{equation}
R(\tau) = R_0 + c_T \int_{\tau_0}^\tau d \tau'  \int_{\tau_0}^{\tau'} d \tau'' \frac{p(\tau'')}{\epsilon(\tau'')}
\end{equation}

The initial radius $R_0$ is taken from overlap calculations. This leaves a parameter
$c_T$ determining the strength of transverse acceleration which is also fit to
data. The final parameter characterizing the expansion is its endpoint given by $\tau_f$,
the breakup proper time of the system. 

\subsection{Thermodynamics}

We assume that entropy is conserved throughout the thermalized expansion phase. Therefore, 
we start by fixing the entropy per baryon from the number of produced
particles per unit rapidity (see e.g. \cite{ENTROPY-BARYON}). 
Calculating the number of participant baryons (see \cite{Dileptons})
we find the total entropy $S_0$. The entropy density at a given proper time
is then determined by $s=S_0/V(\tau)$.

We describe the EoS in the partonic phase by a quasiparticle
interpretation of lattice data which has been shown to reproduce lattice
results both at vanishing baryochemical potential $\mu_B$ and
finite $\mu_B$ \cite{Quasiparticles} (see these references for details of the model).

In the hadron gas phase,
we calculate thermodynamic properties at kinetic decoupling where interactions cease to be important. Here, we 
have reason to expect that an ideal gas will be a good description. Determining
the EoS at this point, we choose a smooth interpolation from here  to the EoS
obtained in the quasiparticle description. This is described in greater detail in
\cite{Dileptons, Thesis}.

With the help of the EoS and $s(\tau)$, we are now in a position to compute
the parameters $p(\tau), \epsilon(\tau), T(\tau)$ as well.
Since the ratio $p(\tau)/\epsilon(\tau)$ appear in the expansion
parametrization, we have to solve the model self-consistently.

\subsection{Solving the model}

In order to adjust the model parameters, we compare with data on transverse momentum
spectra and HBT correlation measurements. 
In \cite{FREEZE-OUT}, a  similar model is fit to a large set of experimental data,
providing  sets of parameters $T_f, v_{\perp f}, R_f, \eta_f^{front}$ at freeze-out.
Although the present model uses a different (box vs. Gaussian) longitudinal distribution of matter,
we use the parameters from this analysis as a guideline for the transverse dynamics
where this difference should not show up and determine $\eta_f^{front}$ separately.
Specifically, we use the set {\bf b1} from  \cite{FREEZE-OUT} for the transverse
dynamics.

By requiring $R(\tau_f) = R_f$ and $v_T^{front} = v_{\perp f}$ we can determine the
model parameters $c_T$ and $\tau_f$. $c_z$ is fixed by the requirement
$\eta^{front} (\tau_f) = \eta_f^{front}$. The remaining parameter $v_0^{front}$
now determines the volume (and hence temperature) at freeze-out and can be adjusted such 
that $T(\tau_f) = T_f$.

The model for 5\% central 158 AGeV Pb-Pb collisions at SPS is characterized by
the following scales: Initial long. expansion velocity $v_0^{front} = 0.5c$, thermalization
time $\tau_0 = 1$ fm/c, 
initial temperature $T_0 = 305$ MeV, duration of the QGP phase $\tau_{QGP} = 7$ fm/c,
duration of the hadronic phase $\tau_{had} = 9$ fm/c, total lifetime $\tau_f - \tau_0
= 16$ fm/c,  r.m.s radius at freeze-out $R_f^{rms} = 8.55$ fm, transverse expansion
velocity $v_{\perp f} = 0.537 c$.

\subsection{Initial compression and re-expansion}

\label{Compression}

\begin{figure}[htb]
\begin{center}
\epsfig{file=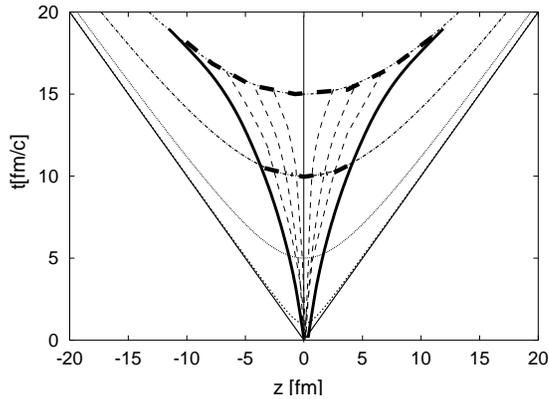, width=7.5cm}
\end{center}
\caption{\label{F-Expansion} A scetch of the accelerated longitudinal expansion pattern
of the fireball matter. 
}
\end{figure}

We have chosen the above model framework in such a way that it allows to account
for the apparent mismatch in timescales derived from Eqs.~(\ref{E-tside}) and (\ref{E-Sinyukov}).
This is done by allowing for initial compression of the matter in longitudinal direction and subsequent
re-expansion. The longitudinal flow pattern of matter is schematically indicated in Fig.~\ref{F-Expansion}.

It is apparent from the figure that a measurement of the conditions around the freeze-out hypersurface 
$\tau = \tau_f$ (where the bulk of hadronic emission takes place)
would not be able to distinguish between an accelerated or non-accelerated
longitudinal evolution history. This has also been pointed out in \cite{FREEZE-OUT}.
Hence, incorporating accelerated longitudinal expansion we have the freedom 
to select $\tau_f$ accoring to Eq.~(\ref{E-tside}) without being in contradiction with
(\ref{E-Sinyukov}). (Since we in practice adjust the model to $R_{long}$ and the
observed multiplicity distribution in rapidity, a detailed discussion of $R_{long}$ in this paper
is not  fruitful).

In our model, the parameter controlling the initial amount of compression is $v_0^{front} = 0.5c$.  
This has profound consequences for the initial state: If we  set $v_0^{front} = v_f^{front}$ (and hence
use the scenario underlying Eq.~(\ref{E-Sinyukov})), the initial temperature of the fireball
drops by $\sim30$\%. 

This in turn has consequences for the emission of real photons.
In \cite{Photons} it was shown that a scenario very similar to the one discussed here 
can account for the measured amount of photon emission, especially 
in the momentum region between 2 and 2.5 GeV where no 
significant prompt photon contribution is expected. Assuming a scenario without longitudinal acceleration,
the yield of direct photons at 2 GeV drops by a factor 8 for the standard choice of the equilibration time
of $\tau_0 \approx 1$ fm/c in striking disagreement with tha data. This cannot fully be compensated even for
a choice of the equilibration time as low as 0.1 fm/c. 

In conclusion, the longitudinally accelerated pattern advocated earlier to explain the difference 
between the timescales extracted from (\ref{E-tside}) and (\ref{E-Sinyukov}) is strongly favoured by 
the direct photon data. Therefore, we will assume in the following that the relevant timescale of the fireball
evolution can be inferred from transverse dynamics only and focus on the remaining question of
the apparent mismatch of the source evolution time $O(15)$ fm/c and its emission duration $O(2)$ fm/c.
In order to investigate this, we have to specify the emission function of the fireball.

\subsection{Hadron emission from the hypersurface with timelike normal}

One contribution to the total emission of particles from the fireball
comes from the final breakup of the thermalized system.
The corresponding freeze-out hypersurface is characterized by a timelike
normal. We calculate particle emission from this phase with the help of
the emission function

\begin{equation}
\begin{split}
\label{E-TimelikeEmission}
S^i_t(x,K) d^4 x  =& \frac{M_T \cosh(Y-\eta_s)}{(2\pi)^3} \exp \left(\frac{K\cdot u(x) + \mu_i}{T_f}
\pm 1 \right)^{-1} \cdot G(r) H(\eta_s) d\eta_s r dr d\phi \\
& \times \frac{\tau d\tau}{\sqrt{2\pi(\Delta\tau)^2}}\exp \left( -\frac{(\tau-\tau_f)^2}{2(\Delta\tau)^2} \right)
\end{split}
\end{equation}

In this expression, the momentum $K$ is parametrized in terms of longitudinal
rapidity $Y$, transverse mass $M_T = \sqrt{K_t^2+m_i^2}$ and transverse momentum $K_t$
The factor $ \exp \left(\frac{k\cdot u(x) + \mu_i}{T_f}
\pm 1 \right)^{-1}$
corresponds to the thermal distribution of particle species $i$ (with corresponding
chemical potential $\mu_i$ and mass $m_i$) where the product $k\cdot u(x)$ takes care of
evaluating the distribution in the rest frame of matter moving with four-velocity $u(x)$
as seen from a particle with momentum $K$ and the $+(-)$ sign holds for fermions (bosons).
$G(r)$ and $H(\eta_s)$ describe the distribution of matter in radial and longitudinal
direction using the radius $r$ and spacetime rapidity $\eta_s$.
Finally, the factor $\exp \left( -(\tau-\tau_f)^2/(2(\Delta\tau)^2) \right)$
can be seen as a smearing of a sudden decoupling occuring at proper time $\tau_f$ into
an extended breakup time period characterized by $\Delta\tau$.

The single particle spectrum for particle species $i$ is calculated from this emission function as

\begin{equation}
\label{E-SingleSpectrumBase}
E^i_p \frac{dN_i}{d^3 p} = \int d^4 x S^i_t(x,p) 
\end{equation}
with $E^i_p = \sqrt{p^2+m_i^2}$.
HBT correlation radii are obtained using the common Cartesian parametrization 

\begin{equation}
C(q,K) -1 = \exp \left[ -q_o^2R_{out}^2(K) - q_s^2R_{side}^2(K)-q_l^2R_{long}^2(K)-2q_oq_lR^2_{ol}(K)\right]
\end{equation}
(see e.g. \cite{HBTReport, HBTBoris} for
an overview and further references)
for the correlator.
Here, $K = \frac{1}{2} (p_1 + p_2)$ is the averaged momentum of the correlated pair 
with individual momenta $p_1,p_2$ and $q = (p_1 - p_2)$ the momentum difference.
The transverse correlation radii $R_{out,side}$ follow from the emission function
as
\begin{equation}
\label{ERSideBase}
R_{side}^2 = \langle \tilde{y}^2 \rangle
\end{equation}
\begin{equation}
\label{EROutBase}
R_{out}^2 = \langle (\tilde{x} - \beta_\perp\tilde{t})^2 \rangle
\end{equation}
with $\beta_\perp$ the transverse velocity of the emitted pair, 
$\tilde x_\mu = x_\mu - \langle x_\mu \rangle$ and
\begin{equation}
\langle f(x) \rangle(K) = \frac{\int d^4 x f(x) S(x,K)}{\int d^4 x S(x,K)}
\end{equation}
an average with the emission function.

In \cite{FREEZE-OUT}, expressions (\ref{E-SingleSpectrumBase}), (\ref{ERSideBase})
and (\ref{EROutBase}) have been used to fit the parameters of the shapes
$G(r)$ and $H(\eta)$, the radial and longitudinal expansion
velocities at freeze-out $\tau_f$, the freeze-out temperature $T_f$  
and the temporal smearing $\Delta\tau$ to experiment, thus
providing the basic parameters for the fireball evolution described above.
As a starting point for the investigations in this paper, 
we assume a box profile $G(r) = \theta(R_B -r)$ for the radial distribution of 
thermalized matter. We use $H(\eta_s) = \theta(\eta_s^{front}(\tau_f) - \eta_s)
\theta(\eta_s^{front}(\tau_f) + \eta_s)$
throughout this paper.

\subsection{Hadron emission from the hypersurface with spacelike normal}

In addition to particle emission at the final breakup of the fireball described by
(\ref{E-TimelikeEmission}),
there is also continuous emission of particles throughout the
whole lifetime of the system. Such emission is characterized by a freeze-out
hypersurface with spacelike normal.

Since the correlation radius $R_{out}$ probes not only the geometrical radius
of the source, but also the duration of particle emission (see Eq.~(\ref{EROutBase})), it may well be that
such a contribution significantly change the ratio $R_{out}/R_{side}$ (which is known 
to be small experimentally) as compared to the use of only the emission function 
(\ref{E-TimelikeEmission}).

In order to investigate this question, we consider in addition to 
(\ref{E-TimelikeEmission}) also the emission function 

\begin{equation}
\label{E-SpacelikeEmission}
\begin{split}
S_s^i(x,K) d^4 x  &= \frac{1}{(2\pi)^3} \left(K_\perp \cos(\phi) - v_\perp(\tau) M_T \cosh(Y-\eta_s)\right)
\cdot \theta\left(K_\perp\cos(\phi) -
v_\perp(\tau) M_T \cosh(Y-\eta_s)\right) \\ & \times \exp \left(\frac{K\cdot u(x) + \mu_i^B}{T_B}
\pm 1 \right)^{-1}  
d\eta_s r dr d\phi \tau d \tau 
\delta(r-R_B(\tau))
\theta(\eta_s + \eta_s^{front}(\tau)) \\ & \times \theta(\eta_s^{front}(\tau)-\eta_s)
\theta(\tau - \tau_i) \theta(\tau_f-\tau)
\end{split}
\end{equation}
which describes the emission of particles with transverse velocity $K_\perp/E_K\cos(\phi)$
from the fireball surface moving at $v_\perp$ (with $\phi$ the angle between surface and
the momentum vector of the emitted particle) at $r=R_B(\tau)$ between the space-time rapidity extent
of the fireball from $+\eta_f(\tau)$ to $-\eta_f(\tau)$ between initial time
$\tau_i$ and freeze-out time $\tau_f$. The factor $\theta\left(\frac{K_\perp}{E_K} \cos(\phi) -
v_\perp(\tau)\right) $ ensures that emission only takes place for particles moving faster than the
emission surface and no negative contribution of 'backward emission' are counted.

We assume that the expanding
fireball can be described as a hot 'core' and a (thin) boundary
region from which the emission takes place. This region is assumed
to be characterized by an average temperature $T_B$. We do not
require $T_B = T_f$, but leave this as a free parameter, however we
do require that the boundary is characterized by the same pion density
as the whole fireball at breakup, which fixes $\mu_i^B$ for given
$T_B$.

To a first approximation, we describe the total particle emission 
by the sum of both emission throughout lifetime and final
breakup  as $S^i(x,k) =S_t^i(x,k) + S_s^i(x,k)$.

\section{Calculation of HBT correlation parameters}

\subsection{$R_{side}$ and the freeze-out geometry}

The experimental results for the correlation radius $R_{side}$ are
often fit by a function of the form

\begin{equation}
\label{E-R_side_simple}
R_{side} = \frac{R_{G}}{\sqrt{1 + m_t \eta_{\perp f}^2/T_f}}.
\end{equation}

This expression comes from an approach where only the final breakup of the
fireball is taken into account, the matter distribution $G(r)$ is assumed to
be Gaussian with width $R_{G}$, and Boltzmann statistics instead of Bose-Einstein statistics is
used.

In contrast, in this paper we study a model with a long duration of the
emission (throghout the complete fireball lifetime of $\sim 17$ fm/c),
and a final breakup, which is characterized rather by a box-shaped transverse
distribution .
In addition, we use the Bose-Einstein (Fermi-Dirac) distribution where appropriate.

Evidently, the analytic expression (\ref{E-R_side_simple}) is only able to determine the
ratio $\eta_{\perp f}^2/T$. Assuming $T_f = 120$ MeV, the fit yields $R_{G}\approx 7$ fm 
corresponding to $R_{rms} \approx 9.8$ fm and
$\overline{ v}_\perp = 0.55 c$ for the most central collisions at 158 AGeV \cite{CERES-HBT}. (Compare with the
parameters in our fireball evolution scenario 
$T_f = 100$ MeV, $\overline{v}_\perp = 0.537 c$ and $R_{rms} =  8.55$ fm.)

\begin{figure}[htb]
\begin{center}
\epsfig{file=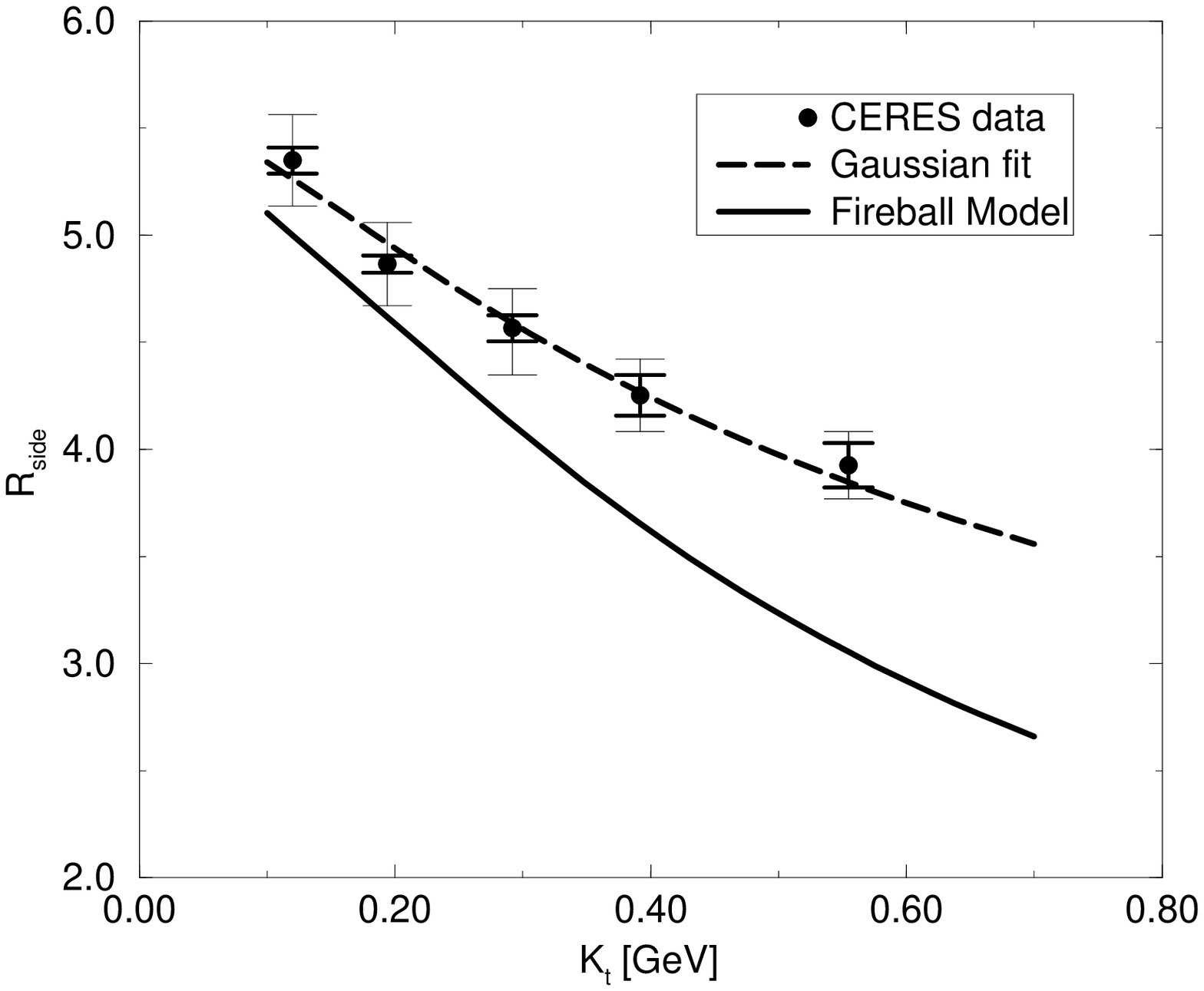, width=7.5cm}\epsfig{file=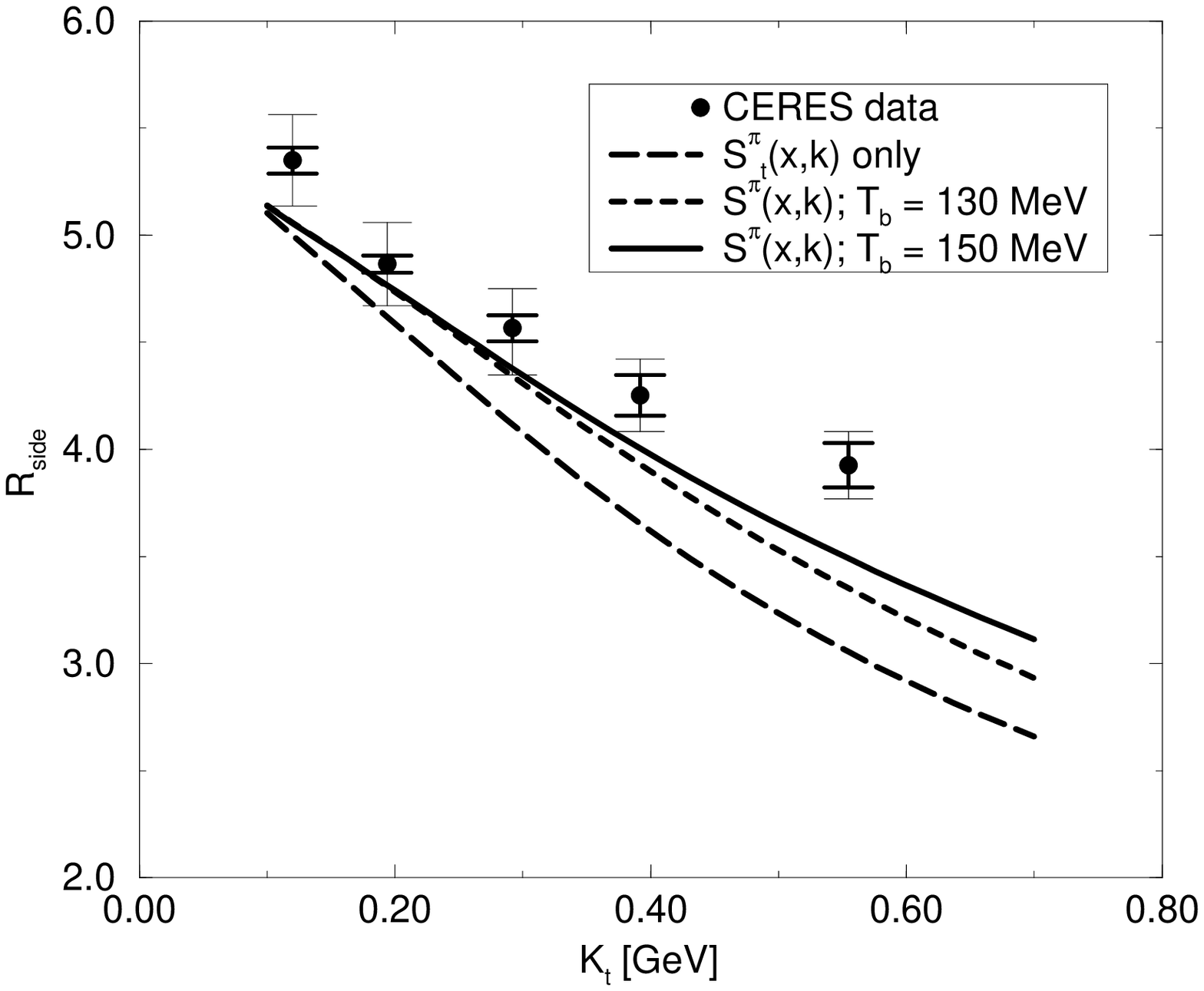, width=7.5cm}
\end{center}
\caption{\label{F-Box_vs_Gauss} Left panel: CERES data on $R_{side}$ for central
Pb-Pb collisions at 158 AGeV with statistical (small bars) and systematical (large bars)
errors (unmerged), in addition to a fit to data based on a Gaussian source and
the result based on the box profile used in our fireball evolution scenario, considering
breakup at $\tau_f$ only. Right panel: Data as before, shown is emission only at fireball
breakup (dashed), and the results with the full emission function for two
different values of the average temperature of the emission layer (dotted and solid). }
\end{figure}

For the sake of clarity, we start by presenting the result of evaluating
(\ref{ERSideBase}) using $S^\pi_t(x,K)$ only, i.e. we neglect all emission
throughout the fireball lifetime. The result is shown in Fig.~\ref{F-Box_vs_Gauss}.

Clearly, the overall normalization (set by the $R_{rms}$ radius) appears too small in
the model and the falloff with transverse pair momentum $K_t$ too steep. 

The latter effect can be explained by
the observation that the Gaussian distribution allows for a (small) component of the total
amount of matter to be at large radii and hence to be boosted with large radial
velocities if a flow profile as in Eq.~(\ref{E-FlowProfile}) is chosen, whereas the
box profile used in the present scenario does not. 

This discrepancy is significantly reduced by using the full emission function
$S^\pi(x,K)$ instead of $S^\pi_t(x,K)$. This is due to the fact that emission
of fast particles is preferred for a freeze-out hypersurface with spacelike normal,
leading to an enhancement of the correlation radii at larger $K_t$ also (Fig.~\ref{F-Box_vs_Gauss}, 
right panel). In addition, we have some freedom to adjust $T_B$ (the average temperature of
the emitting region in $S^\pi_s(x,K)$) within reasonable limits. The data
prefer a rather large value $T_B\sim 150$ MeV. This should not come as a surprise, since
the temperature of the emitting layer may well be large in the initial stages
of fireball evolution. 

Guided by the success of the Gaussian parametrization, we
drop the unphysically sharp cutoff of the box profile $G(r) = \theta(R_{box}-r)$ and use instead
a Woods-Saxon distribution (retaining the value of $T_B = 150$ MeV in the following)

\begin{equation}
G(r) =  \frac{1}{1 +\exp \left[\frac{r - R_{box}}{d_{WS}} \right]}
\end{equation}

\begin{figure}[htb]
\begin{center}
\epsfig{file=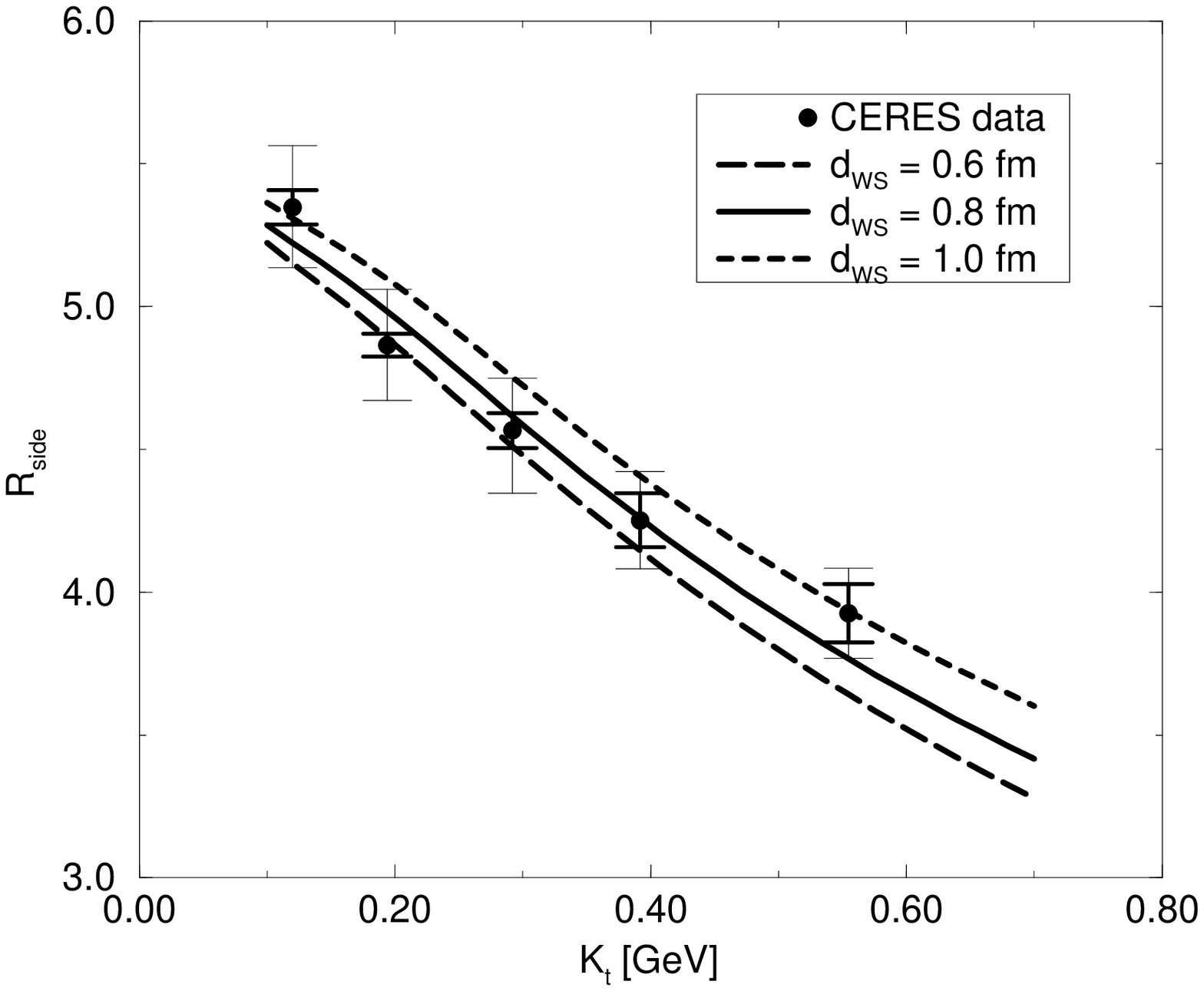, width=7.5cm}\epsfig{file=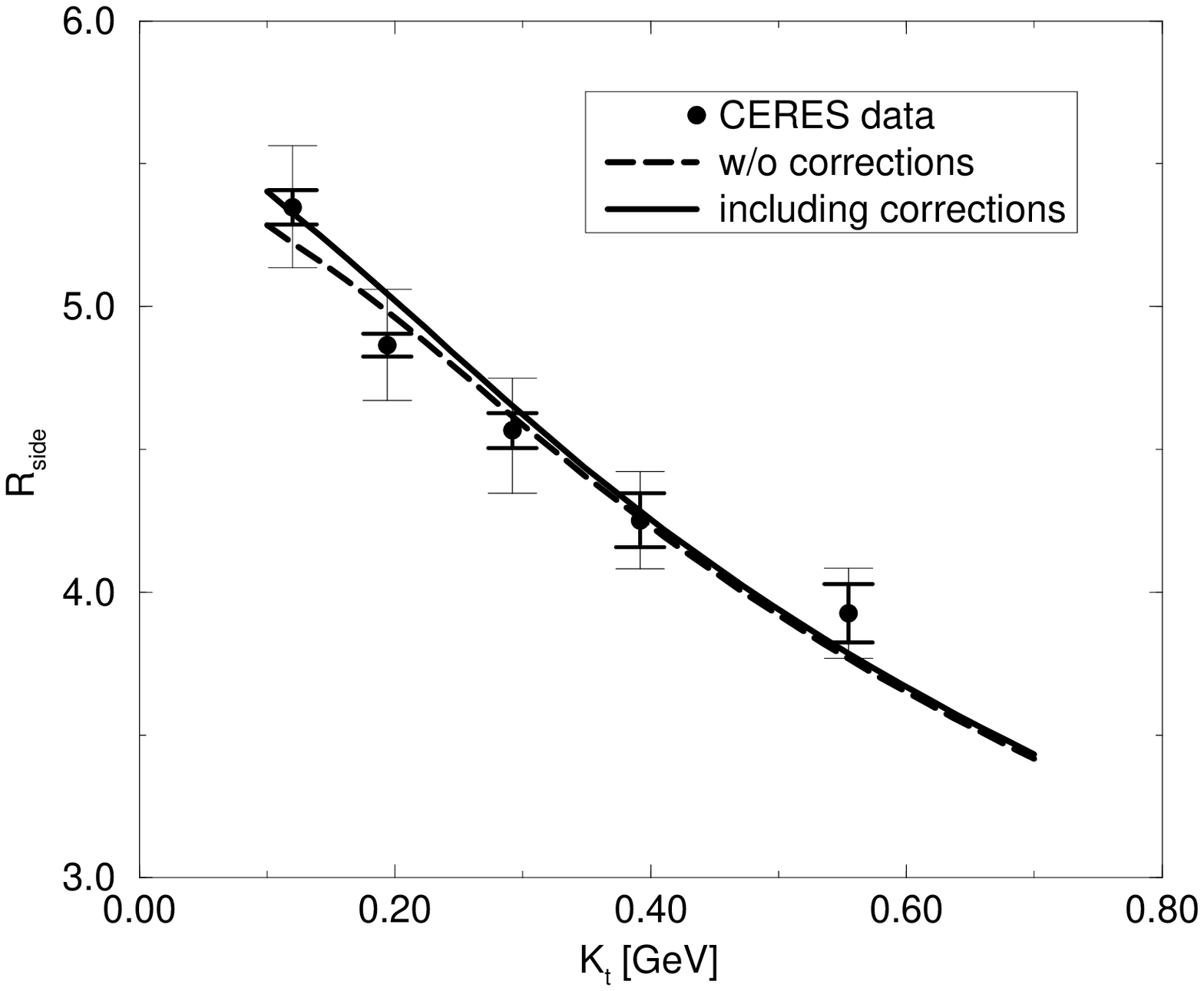, width=7.5cm}
\end{center}
\caption{\label{F-WS} Left panel: CERES data on $R_{side}$ for central
Pb-Pb collisions at 158 AGeV with statistical (small bars) and systematical (large bars)
errors (unmerged). Left panel: Calaulated $R_{side}$ for various values of
the Woods-Saxon skin thickness parameter $d_{WS}$. Right panel: Calculated
$R_{side}$ using $d_{WS} =0.8$ fm (dashed), shown is the effect of model-independent
correction terms (\ref{E-Corrections}) (solid). }
\end{figure}

We find that the data prefer rather small values of $d_{WS}$ of
the order of $0.8$ fm (see Fig. \ref{F-WS}, left panel). This is reassuring since
it gives some justification for the use of a box profile in the calculation
of dilepton and photon emission. 

Finally, we also calculate the model-independent corrections \cite{Heinz-MI} to (\ref{ERSideBase})

\begin{equation}
\label{E-Corrections}
\Delta R_{side}^2 = \left(\frac{1}{4K_\perp}\frac{d}{d K_\perp}
- \frac{1}{2}\frac{d}{dm^2} \right) \ln \int d^4 x S^\pi(x,K),
\end{equation}
which primarily affect the low momentum region. This is shown in Fig.~\ref{F-WS}, right panel.
These corrections are small, however, they tend to improve agreement with data
slightly in the low $K_t$ region.

In summary, including the emission throughout the fireball lifetime and all correction terms,
we achieve a good description of the experimental data.

\subsection{$R_{out}$ and the freeze-out criterion}

It is commonly deduced from (\ref{ERSideBase}) and (\ref{EROutBase}) that
the duration of pion emission can be calculated as
\begin{equation}
\Delta\tau^2 = \frac{1}{\beta_t^2} (R_{out}^2 - R_{side}^2) 
\end{equation}
with $\beta_t$ the velocity of the emitted pair $\beta_t = K_t/E_K$.
However, there are in fact two different timescales related to the duration of
the emission in our model: a) the total lifetime of the fireball of order $\sim 17$ fm/c
and b) the duration of the final breakup, which is a priori unknown, but can be
expected to be of order of at most a few fm/c for the whole evolution scenario to
make sense. We therefore expect the actual difference between $R_{out}$ and
$R_{side}$ in our model to be characterized by a mixture of these two
scales, weighted with their respective contribution to the total pion emission.

\begin{figure}[htb]
\begin{center}
\epsfig{file=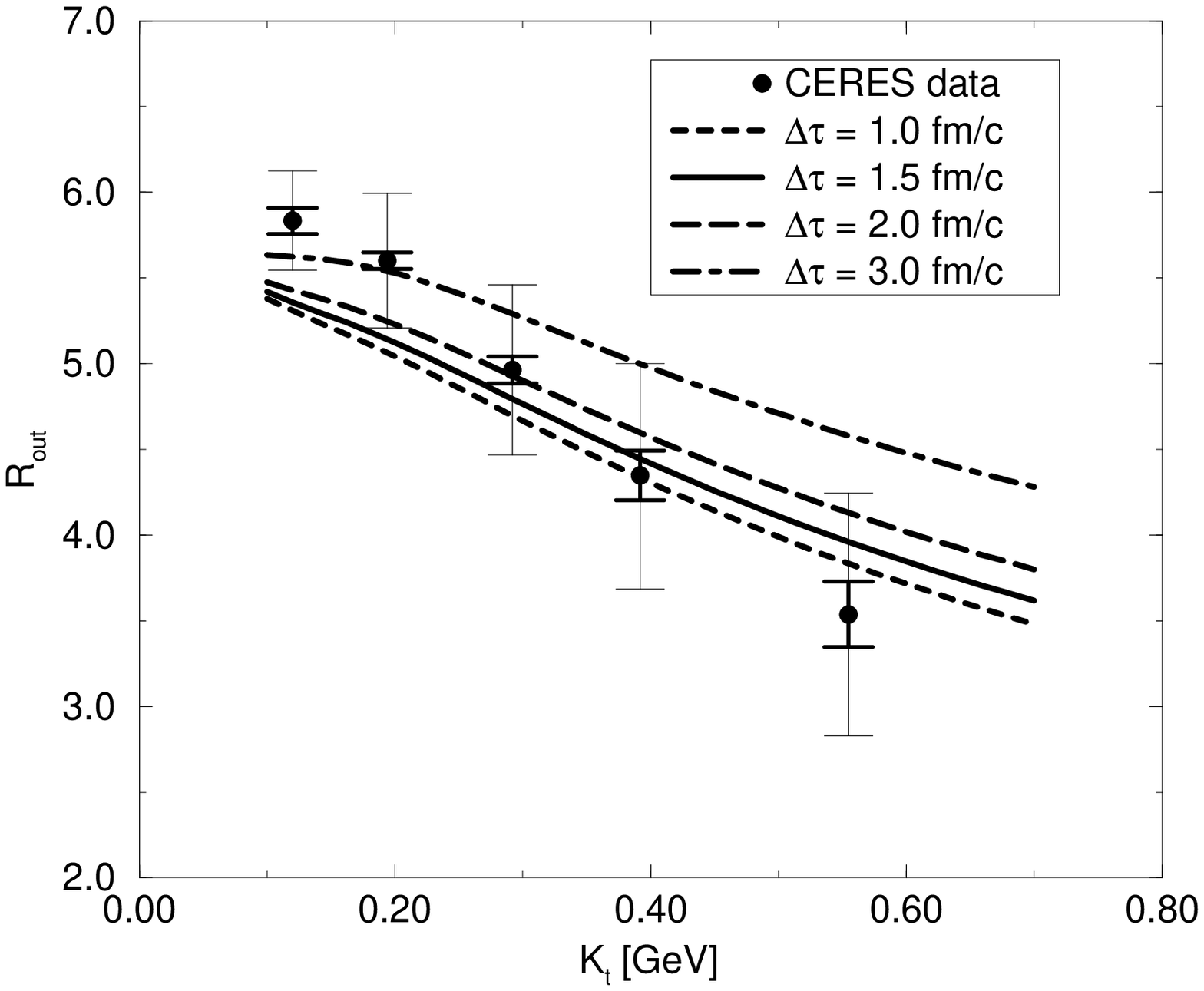, width=7.5cm}\epsfig{file=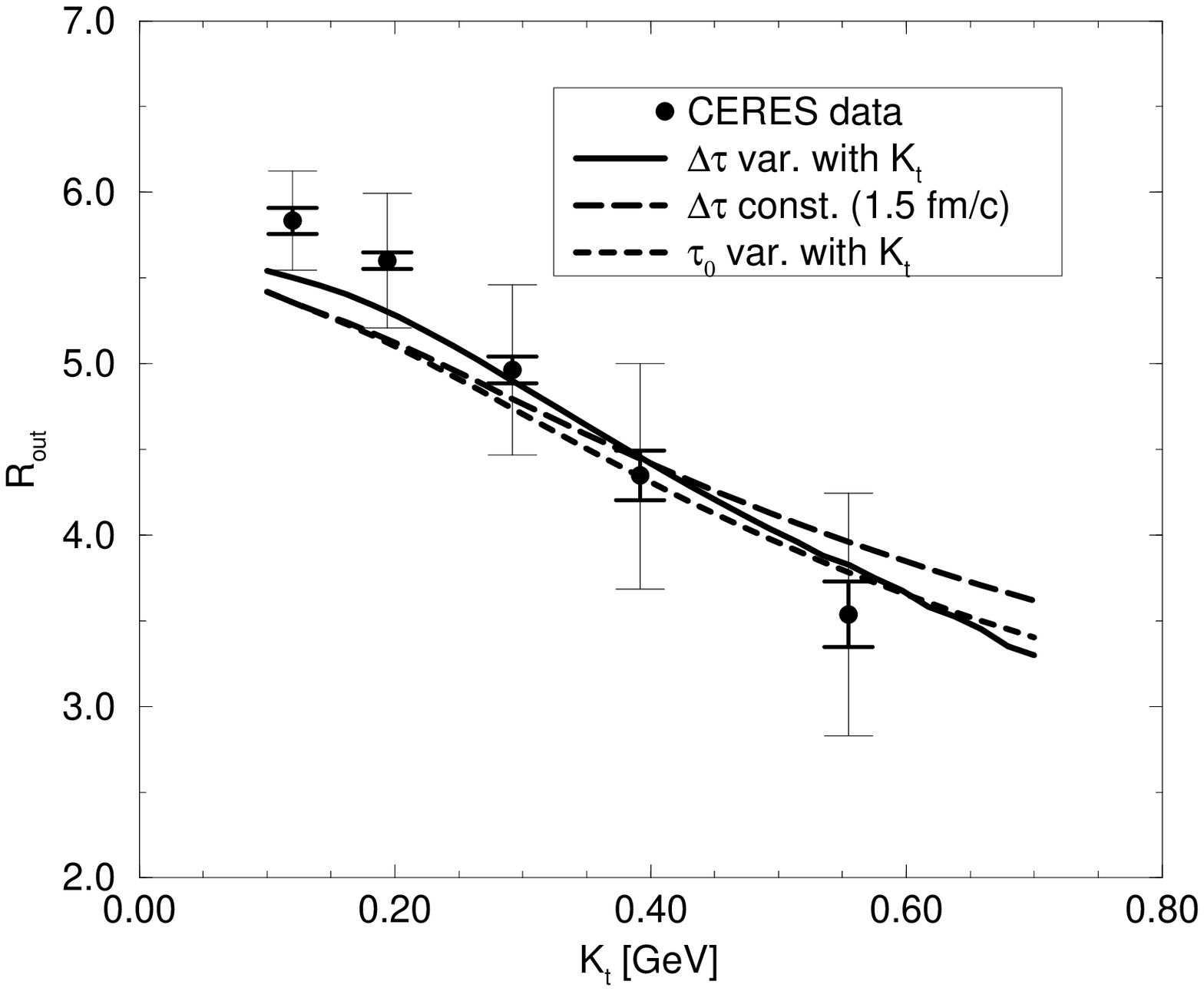, width=7.5cm}
\end{center}
\caption{\label{F-R_out_dt} CERES data on $R_{out}$ for central
Pb-Pb collisions at 158 AGeV with statistical (small bars) and systematical (large bars)
errors (unmerged). Left panel: Calculated $R_{out}$ for different values of
the fireball breakup time $\Delta\tau$. Right panel: $R_{out}$ for
different momentum-dependent freeze-out scenarios (see main text)}
\end{figure}

We start the investigation of $R_{out}$ by fixing all parameters at the
value used to describe $R_{side}$ and evaluating (\ref{EROutBase})
with different values of $\Delta \tau$, the final breakup duration.
In all following calculations, we also include the correction term for
$R_{out}$ \cite{Heinz-MI}
\begin{equation}
\Delta R^2_{out} = \left(\frac{1}{4}\frac{d^2}{dK_t^2} - \frac{1}{2}
(1-\beta_\perp^2)\frac{d}{dm^2} \right) \ln \int d^4x S^\pi(x,K)
\end{equation}
The result is shown in Fig.~\ref{F-R_out_dt}.

Clearly, the overall magnitude of the calculated $R_{out}$ is in agreement with
data, provided that a rather sudden breakup time of the order of 1--3 fm/c is
choosen. This indicates that that the difference $1/\beta_t^2 (R^2_{out} - R^2_{side})$ does not
reflect the full fireball lifetime $\tau_f^2$ but rather is dominantly influenced
by $\Delta\tau$ only (otherwise one would expect the difference to be significantly
larger). We will argue later on that this is so because the numerical contribtion
from continuous emission throughout the fireball lifetime is small compared to
the contribution from final breakup.

The shape of the falloff with $K_t$, however, is not well reproduced for any choice
of $\Delta\tau$. (In fact, the shape appears similar to the falloff of
$R_{side}$ and rather insensitive to variations of $\Delta\tau$.)
This suggests that the data (albeit the systematic errors are rather large)
indicate the need for a more refined freeze-out scenario.

In \cite{Wiedemann-FO}, such a more detailed scenario based on the
(momentum-dependent) mean free path of pions inside the hot matter
is described. It is beyond the scope of the present paper to
show that this freeze-out criterion is able to describe the
data, however, we will argue that it leads to the correct qualitative
behaviour of the calculated $R_{out}$. 

The essential observation made in \cite{Wiedemann-FO} ist that
the freeze-out is different for high momentum pions than for
low momentum pions: High momentum pions are expected to escape
earlier. 

If we want to incorporate this into our calculation, we have to
make either $\Delta\tau$ or $\tau_f$ (or both) dependent on
$K_t$, the idea being that the final breakup of the fireball
occurs earlier for high momentum pions and happens faster.

In the right panel of Fig.~\ref{F-R_out_dt}, we demonstrate that
the dependence of $\Delta\tau$ is able to qualitatively
improve the agreement with data. To obtain an estimate
for the qualitative behaviour, we assume $\Delta\tau = 2.0$ fm/c
for $K_T = 100$ MeV and $\Delta\tau = 1.0$ fm/c for $K_t = 700$ MeV
and interpolate linearly inbetween. In view of the large systematic
uncertainties, however, we do not push this idea further by trying to aim
at a perfect description of the data. 

As for the dependence of $\tau_f$ on $K_t$,
we observe that earlier freeze-out leads in general to smaller
$R_{out}$ at larger $K_t$, since one probes a radially less expanded
system. A suitable variation of $\Delta\tau$ might then be able to
lead to a good description of the data. Again, in view of the
large uncertainties we refrain from a further examination
of this scenario.

\section{Pion emission and fireball thermodynamics}

The main result of the previous section is that $R_{out}$ does
not seem to reflect the order of magnitude of the fireball lifetime $\tau_f$, even if
continuous emission of pions is taken into account. This is
a rather surprising result, and in this section we will investigate
the relevant emission function $S_s^\pi(x,k)$ further in order to
answer the following questions:

\begin{itemize}
\item
Why is $\Delta\tau$ and not $\tau_f$ the dominant timescale in $R_{out}$?
\item
Does the continuous loss of pions (and nucleons) have any effect on the
thermodynamics of the fireball (and hence invalidate the findings of e.g. \cite{Dileptons})?
\end{itemize}

In order to answer these questions, we investigate the transverse velocity distributions
in the c.m. frame of pions and nucleons emitted before $\tau_f$ (which can be easily calculated by integrating
(\ref{E-SpacelikeEmission}) over the fireball lifetime). The reason for choosing this particular
distribution is the following: We may well imagine a process where a
particle is emitted in the initial stage of the expansion. Such particle
can be emitted with relatively small velocity (as seen in the c.m. frame), since
the fireball initially does not expand in radial direction. If the emission
velocity of such a particle is, however, smaller than the radial
velocity of the fireball at $\tau_f$, there is a chance that this already
emitted particle re-enters the thermalized region. The condition reads:
\begin{equation}
\label{E-Recapture}
R_B(t_E) + v_E^\perp t = R_B(t_E + t); t_E + t < t_f
\end{equation}
with emission time $t_E$ and emission velocity $v_E$ measured in the c.m. frame.
This condition is most easily implemented using the transverse
velocity distribution of emitted particles.

\begin{figure}[htb]
\begin{center}
\epsfig{file=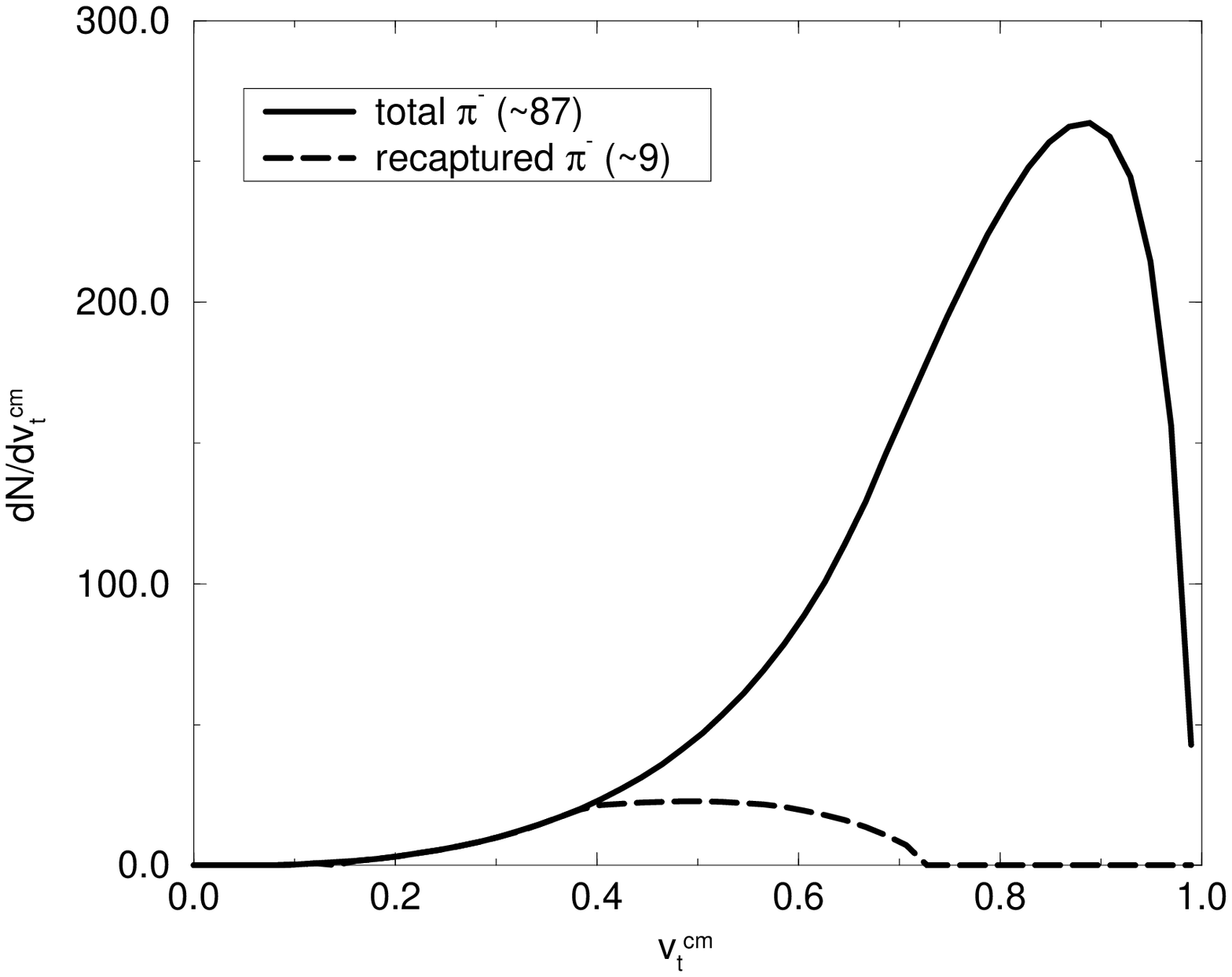, width=7.5cm}\epsfig{file=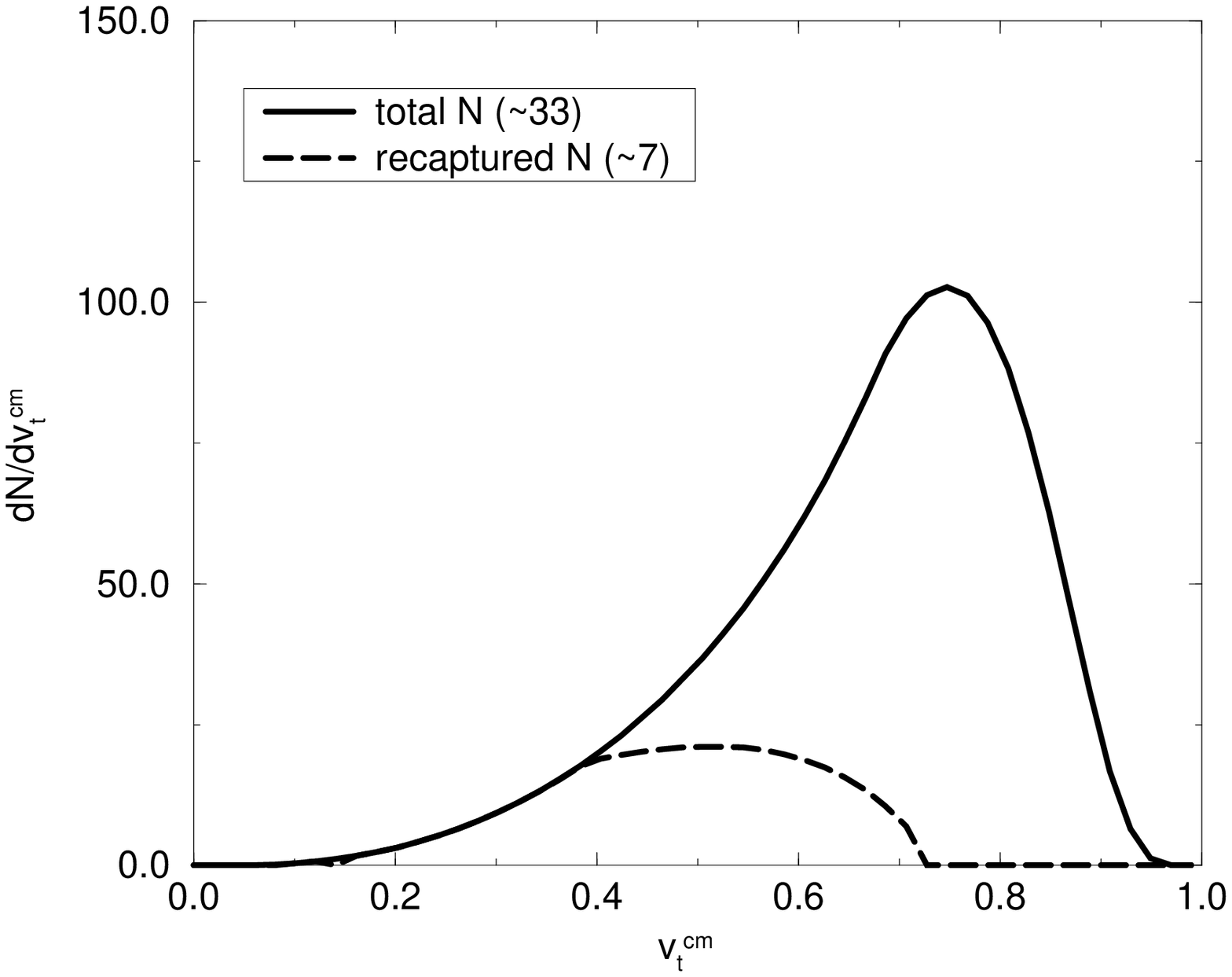, width=7.5cm}
\end{center}
\caption{\label{F-dNdv}Left panel: Transverse velocity distribution of
total $\pi^-$ emission and of recaptured $\pi^-$. Right panel: Same for nucleons.
Also given is the integral over the distribution. }
\end{figure}

The resulting transverse velocity distributions are shown in Fig.~\ref{F-dNdv}.

Several important observations can be made: For the emission of pions,
the recapturing process is not very efficient. This is due to the fact that
the mass of the pions is of order of the temperature of the emitting layer,
hence the transverse velocity distribution is peaked towards large velocities
where recapture cannot take place. On the other hand, the total number of
pions emitted before the final breakup of the fireball is small as compared to
the total number of observed $h^-$ ($\sim 700$). This explains partially the
observed insensitivity of $R_{out}$ to the fireball lifetime. 

Turning to nucleons, we observe that the recapturing process is somewhat more
efficient due to the large nucleon mass leading to reduced emission velocities.
The total number of emitted nucleons, however, is also small as compared to
the total number of participants ($\sim 350$). 

Note in addition that most of the emission takes place close to $\tau_f$ 
anyway, since the area of the emitting surface grows strongly with time as
the fireball expands and the total number of emitted particles per unit time is
proportional to this area. This is enhanced by the fact that
the fireball in this scenario undergoes accelerated expansion in both in longitudinal and
transverse direction, thus the emitting surface is for the larger part of the
evolution significantly smaller than its final value.

We can incorporate the early loss of pions schematically into
the calculation of dilepton emission (see \cite{Dileptons})(the observable which is
presumably most affected by this effect) by lowering the value of
the pion chemical potential $\mu_\pi$ during the hadronic evolution stage
(in \cite{Dileptons}, we assumed a linear rise with proper time
of $\mu_\pi$ from 0 at the phase transition to a value consistent
with the observed number of pions at thermal freeze-out. 
Now, we choose the endpoint such that the number of pions from both 
continuous emission and final breakup agrees with the observed number. We
find $\mu_\pi(\tau_f) \approx$ 110 MeV).

The calculation reveals that the dilepton yield in the low invariant mass
region between 300 and 700 MeV is reduced by $\sim 10\%$ by the
correction (this is hardly visible in a logatithmic plot, therefore we do
not show a graph). Above 1 GeV invariant mass, the dilepton
yield is dominated by Drell-Yan pairs and emission from the hot
QGP, consequently this domain remains almost unaltered be the correction.

In summary, we conclude that it is sufficient for
the description of gross thermodynamical properties of the expanding
matter to approximate the fireball as a closed system which does not
emit strongly interacting partcles before $\tau_f$.

\section{Conclusions}

We have investigated the effect of particle emission prior to the
fireball breakup time $\tau_f$ in a framework
based on gross thermodynamic properties of hot QCD matter, which is
nevertheless able to account for a large number of experimental
observables and discussed the connection to the  
HBT parameters $R_{out}$ and $R_{side}$.

Calculating the emission of pions from the fireball boundary and breakup,
we computed
$R_{out}$ and $R_{side}$ and compared to measurements by the
CERES collaboration. We found that $R_{side}$ can be well
decribed within the present framework.

Regarding $R_{out}$, we identified two timescales which
potentially enter this quantity, the fireball breakup time
$\Delta\tau$ and the total lifetime $\tau_f$. However, in the
present framework, only $\Delta\tau$ was shown to significantly 
influence $R_{out}$. In addition, we found that a momentum-independent
freeze-out criterion does not provide the best description
of the data. We argued qualitatively that a more refined
criterion (e.g. as suggested in \cite{Wiedemann-FO}) improves the
agreement with data.

Finally, we demonstrated that the scenario discussed here does not
invalidate our previous results where we did not include the
effects of continuous emission of particles from the thermalized matter.

\section*{Acknowledgements}

I would like to thank W.~Weise, H.~Appelsh\"{a}user and A.~Polleri, 
for interesting discussions, helpful comments and support.

\end{document}